\documentclass[twocolumn,showpacs,preprintnumbers,amsmath,amssymb,aps,pra]{revtex4-1}
\usepackage{lipsum}
\usepackage{graphicx}
\usepackage{epstopdf}
\usepackage{dcolumn}
\usepackage{bm}
\usepackage{float}
\usepackage{amsmath}
\usepackage{subfig}
\begin{document}     
      

\title{ Hamilton's Dynamics in Complex Phase Space}

\author{M. A. Shahzad}
 \affiliation{ Department of Physics, Hazara University, Pakistan.}
\email{Email:muhammad.shahzad@unicam.it}

\date{\today}

\begin{abstract}
We present the basic formulation  of Hamilton dynamics in complex phase space.
We extend the  Hamilton's function by including the imaginary part and find out the corresponding Hamilton's canonical equation of motion. Example of simple harmonic motion are considered and the corresponding trajectory are plotted on real and complex phase space. 
\end{abstract}


\maketitle

Let $H(q,p,t)=U(q,p,t)+jV(q,p,t)$ be a complex Hamilton's function \cite{H1,H2,H3} in complex phase space, where $q$ is the generalized coordinate and $p$ the conjugate generalized momentum defined on the  phase space $(q,p)$. The total differentiation of the Hamiltonian $H(q,p,t)$ is given by
\begin{align}
dH={\partial U\over \partial q}dq+j{\partial V\over \partial q}dq+{\partial U\over \partial p}dp+j{\partial V\over \partial p}dp +{\partial H\over \partial t}dt
\end{align}
The relation of  Lagrangian function $L(q,\dot{q},t)$ with Hamilton's function is defined by the following equation
\begin{equation}
H=p\dot{q}-L(q,\dot{q},t)=p\dot{q}-\big[W(q,\dot{q},t)+jZ(q,\dot{q},t)\big]
\end{equation}
Taking total differentiation of equation (2), we obtain
\begin{align}
dH=pd\dot{q}+\dot{q}dp-{\partial L\over \partial \dot{q}}dq-i{\partial L\over \partial q}dq-{\partial L\over \partial t}dt
\end{align}
Defining the generalized momentum conjugate to $q$ as
\begin{equation}
p={\partial L\over \partial \dot{q}}
\end{equation}
and using 
\begin{equation}
{\partial L\over \partial q}d q={d\over dt}{\partial L\over \partial\dot{q}}dq=\dot{p}dq
\end{equation}
we obtain
\begin{equation}
dH=\dot{q}dp-\dot{p}dq-\big(\partial L/\partial t\big)dt
\end{equation}
Since, $q$ and $p$ are independent variables, the variations of $dq$, $dp$ and $dt$ are mutually independent. As a result their coefficients must be equal in Equation (1) and (6). Hence
\begin{equation}
\dot{q}={\partial U\over \partial p}+j{\partial V\over \partial p}
\end{equation}
\begin{equation}
\dot{p}=-{\partial U\over \partial q}-j{\partial V\over \partial q}
\end{equation}
and 
\begin{equation}
{\partial H\over \partial t}=-{\partial L\over \partial t}
\end{equation}
\begin{figure}[htbp]
\subfloat[]{\includegraphics[scale=0.565]{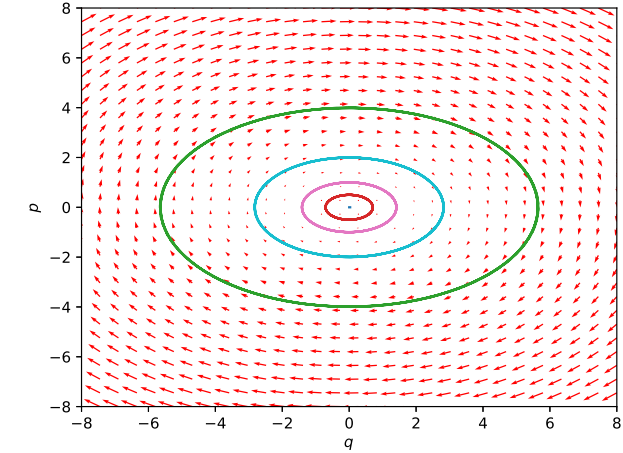}}\\
\subfloat[]{\includegraphics[scale=0.565]{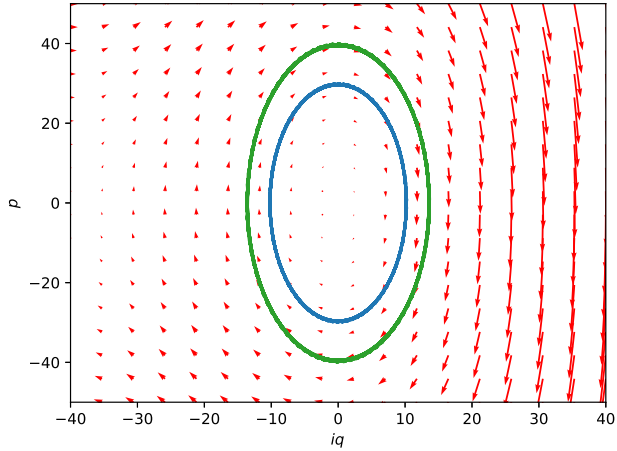}}\\
\subfloat[]{\includegraphics[scale=0.565]{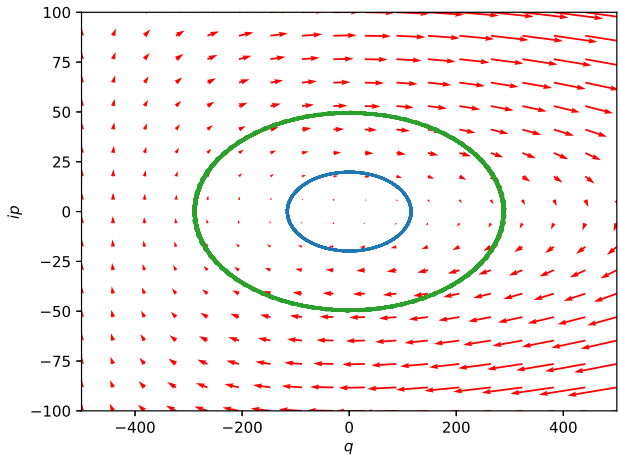}}\\
\caption{ Trajectories of simple harmonic motion on phase space $(q,p)$, $(jq,p)$ and $(q,jp)$.\label{LBP}}
\label{fig:digraph}
\end{figure}
These first-order differential equations are called as Hamilton's canonical equation of motion. By considering a complex Hamilton function of the form $H(q,p,t)=U(q,p,t)+jV(q,p,t)$, we obtained the Hamilton's canonical equation of motion for the Hamilton's function $H(q,p,t)$ defined on complex phase space.
In order to understand the Hamilton dynamics in complex phase-space, we consider an example of simple harmonic motion defined by the following  Hamilton's function $H(q,p,t)$ with imaginary part equal to zero,
\begin{equation}
H(q,p,t)=U(q,p,t)+jV(q,p,t)={P^2\over 2m}+ {k\over 2} q^2\nonumber
\end{equation}
From equation (7,8,9), we have
\begin{align}
\dot{q}&={\partial H\over \partial p}={p\over m}\qquad \dot{p}=-{\partial H\over \partial q}=-kq\nonumber\\
&{\partial H\over \partial t}=0\nonumber
\end{align}
Figure (1(a)) shown the phase space trajectory of simple harmonic motion. \\
Consider an example of simple harmonic motion defined by the following complex Hamilton's function $H(q,p,t)$ with non-zero imaginary part,
\begin{equation}
H(q,p,t)=U(q,p,t)+jV(q,p,t)={P^2\over 2m}+ j{k\over 2} q^2\nonumber
\end{equation}
The corresponding equation of motion can be written as
\begin{align}
\dot{q}&={\partial H\over \partial p}={p\over m}\qquad \dot{p}=-{\partial H\over \partial q}=-jkq\nonumber\\
&{\partial H\over \partial t}=0\nonumber
\end{align}
Figure (1(b),1(c)) shown the trajectory of simple harmonic motion on complex phase space.

Let $F(q,p)$ and $H(q,p)$, with $H(q,p)=U(q,p)+jV(q,p)$,  be two functions on phase space. Then the Poisson bracket is given by
\begin{equation}
[F,H]=[F,U]+j[F,V]
\end{equation}
In particular, $[q_i,q_i+jp_k]=[q_i,q_i]+j[q_i,p_k]=j\delta_{ik}$ and  $[q_i,jp_k]=[q,0]+j[q,p]=j\delta_{ik}$. 
With the help of Hamilton's equations (7,8,9), we have
\begin{equation}
{dF\over dt}=[F,H]+{\partial F\over \partial t}=[F,U]+j[F,V]+{\partial F\over \partial t}\nonumber
\end{equation}

Consider a  bivariate  Hamilton function $H'(q,p)$ associated to the univariate Hamilton function $H(z)$
via $H'(x, y) = U(x, y) + jV (x, y) = H(z)|_{z=x+jy}$. The total differential is  defined as \cite{P}
\begin{equation}
dH'={\partial H'(q,p)\over \partial q}dq+{\partial H'(q,p)\over \partial p}dp\nonumber
\end{equation}
Using $H'(q,p)=U(q,p)+jV(q,p)$, we obtain
\begin{eqnarray}
dH'={\partial U(q,p)\over \partial q}dq+j{\partial  V(q,p)\over\partial q}dq\nonumber\\
+{\partial U(q,p)\over\partial p}dp+j{\partial U(q,p)\over\partial p}dp\nonumber
\end{eqnarray}
Since, $z=q+jp$ and $z^{*}=q-jp$, we have
\begin{eqnarray}
dz=dq+jdp\nonumber\\
dz^{*}=dq-jdp\nonumber
\end{eqnarray}
and 
\begin{eqnarray}
dq={1\over 2}(dz+dz^{*})\nonumber\\
dp={1\over 2j}(dz-dz^{*})\nonumber
\end{eqnarray}
The total differential $dH$ of a complex valued Hamilton function $H(z)$ can be expressed as
\begin{equation}
dH={\partial H(z)\over \partial z}dz+{\partial H(z)\over \partial z^{*}}dz^{*}
\end{equation}
where the Wirtinger derivatives \cite{W} are defined by
\begin{eqnarray}
{\partial \over \partial z}={1\over 2}\Big[{\partial\over \partial q}-j{\partial \over \partial p}\Big]\\
{\partial \over \partial z^{*}}={1\over 2}\Big[{\partial\over \partial q}+j{\partial \over \partial p}\Big]
\end{eqnarray}
Using $H=p\dot{q}-L$, we have
\begin{align}
dH=pd\dot{q}+\dot{q}dp-{\partial L\over \partial z^1}dz^1-{\partial L\over \partial z^{1*}}d z^{1*}
\end{align}
where
\begin{align}
z^1&=(q+j\dot{q})\nonumber\\
z^{1*}&=(q-j\dot{q})\nonumber\\
dq&={1\over 2}(dz^1+dz^{1*})\nonumber\\
d\dot{q}&={1\over 2j}(dz^1-dz^{1*})\nonumber
\end{align}
Equation (14) becomes
\begin{eqnarray}
dH=pd\dot{q}+\dot{q}dp-{1\over 2}\Big( {\partial\over \partial q} -j{\partial\over \partial \dot{q}}     \Big)L \big(dq+jd\dot{q}\big)\nonumber\\
-{1\over 2}\Big( {\partial\over \partial q} +j{\partial\over \partial \dot{q}} \Big)L \big(dq-jd\dot{q}\big)
\end{eqnarray}
Using equation (4) and (5), equation (15) can be rewritten as
\begin{align}
dH=\dot{q}dp-\dot{p}dq
\end{align}
Using $dq=(dz+dz^{*})/2$, and  $dp=(dz-dz^{*})/2j$, we have
\begin{align}
dH&={\dot{q}\over 2j}(dz-dz^*)-{\dot{p}\over 2}(dz+dz^{*})
\end{align}
or
\begin{align}
dH=\Big({\dot{q}\over 2j}-{\dot{p}\over 2}\Big)dz-\Big({\dot{q}\over 2j}+{\dot{p}\over 2}\Big)dz^*
\end{align}
Comparing the  coefficient of $dz$ and $dz^*$ in equation (18) with equation (11), we obtain
\begin{eqnarray}
\dot{q}=j\Big({\partial \over \partial z}-{\partial\over\partial z^*}\Big)H\\
\dot{p}=-\Big({\partial \over \partial z}+{\partial\over\partial z^*}\Big)H
\end{eqnarray}
Equation (19-20) are Hamilton equation of motion in complex phase space. Substituting back the  Wirtinger derivatives (equation (12,13)) in equation (19-20), we obtain Hamilton equation of motion in real phase space;
\begin{eqnarray}
\dot{q}={\partial H \over \partial p}\\
\dot{p}=-{\partial H\over \partial q}
\end{eqnarray}
Equation (19-20) are basic Hamilton's canonical  equation of motion which can be used to understand the dynamics of particles in complex phase space.



\end{document}